# Smoke Sky - Exploring New Frontiers of Unmanned Aerial Systems for Wildland Fire Science and Applications

16 October 2018

# Executive Summary

**Team:** E. Natasha Stavros (398P), Ali Agha (347F), Allen Sirota (347A), Marco Quadrelli (3471), Kamak Ebadi (347F), Kyongsik Yun (349B)
Jet Propulsion Laboratory, California Institute of Technology

**Collaborators:** US Interagency UAS Advisory Group for Fire Management Joaquin Ramirez, Tecnosylva

**Acknowledgement:** The research was carried out at the Jet Propulsion Laboratory, California Institute of Technology, under a contract with the National Aeronautics and Space Administration

Wildfire has had increasing impacts on society as the climate changes and the wildland urban interface grows. As such, there is a demand for innovative solutions to help manage fire. Managing wildfire can include proactive fire management such as prescribed burning within constrained areas or advancements for reactive fire management (e.g., fire suppression). Because of the growing societal impact, the JPL BlueSky program sought to assess the current state of fire management and technology and determine areas with high return on investment.

To accomplish this, we met with the national interagency Unmanned Aerial System (UAS) Advisory Group (UASAG) and with leading technology transfer experts for fire science and management applications. We provide an overview of the current state as well as an analysis of the impact, maturity and feasibility of integrating different technologies that can be developed by JPL.

Based on the findings, the highest return on investment technologies for fire management are first to develop single micro-aerial vehicle (MAV) autonomy, autonomous sensing over fire, and the associated data and information system for active fire local environment mapping. Once this is completed for a single MAV, expanding the work to include many in a swarm would require further investment of distributed MAV autonomy and MAV swarm mechanics, but could greatly expand the breadth of application over large fires. Important to investing in these technologies will be in developing collaborations with the key influencers and champions for using UAS technology in fire management.

### FINDINGS
- Fire plays an important role in the Earth System carbon cycle and climate system
- Wildfire management has become an issue of national priority as marked by the congressional 2018 H.R. 2862, the Wildfire Disaster Funding Act
- The highest return on investment technologies are autonomy, autonomous sensing over fire, and the associated data and information system
- Technologies can be developed for single UAS, but the full benefit comes from a swarm of UAS
- With small to moderate scale investment there is a lot of technology that can be developed for fire management that is synergistic with JPL investments
- There are a number of domestic and international organizations working to develop technology (and UAS) for fire management, but none offer the expertise that JPL has
- A step beyond single aerial vehicles and the technology of swarm would further enhance fire management

### RECCOMENDATIONS
- Develop single micro-aerial vehicle (MAV) autonomy, autonomous sensing over fire, and the associated data and information system for active fire local environment mapping
- Develop these technologies with CalTech CAST and outside (domestic and international) organizations to ensure utility of the technology for fire science and management applications

# 1 Significance of Wildfire

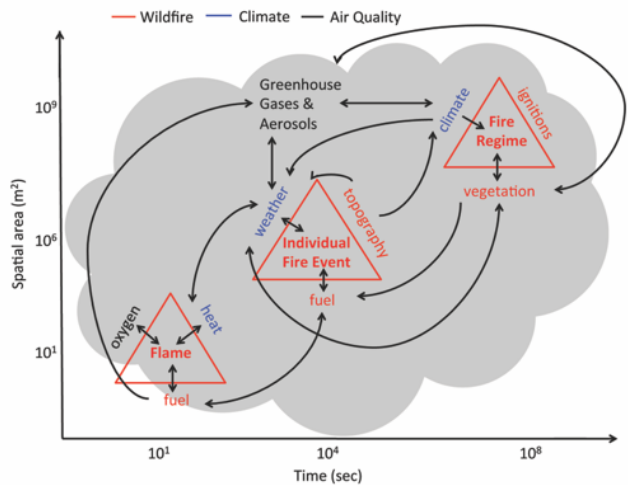

Figure 1: The spatial and temporal resolutions of feedbacks in the wildfire- climate- air quality system. Here air quality is defined to include pollutants and greenhouse gases (Stavros et al. 2014)

Wildfire and its impacts are a complex system with feedbacks and interactions across spatial and temporal scales that affect the Earth System (Figure 1). Specifically, biomass burning is the second largest source of trace gases (Crutzen and Andreae, 1990) and the largest global source of primary, fine carbonaceous particles (Bond et al., 2004), thus substantially contributes to the global carbon cycle (Page et al., 2002). Because of the significance that fire plays in the Earth System, changes in fire area burned, frequency, or severity can have substantial impacts.

In recent years, the number of wildfires, acreage burned, and the length of the fire season in the United States has increased (Krawchuk et al., 2009; Littell et al., 2009; Westerling, 2006), with an increasing trend in megafires (Stavros et al., 2014a). Megafires are defined as fires that have the most substantial impacts and extreme fire behavior (e.g., growing rapidly out of control). Conflating climate-driven changes in wildfire, the wildland urban interface is expanding (Hammer et al., 2009), thus increasing the societal impact from these fire. From 1995 to 2015, US wildfires resulted in $21b (USD 2015) of property losses (NatCatSERVICE), and an increase in the US Forest Service fire response spending from 16% to 52% of its total budget (USFS, 2015). Impacts are not only direct (e.g., loss of life, structures, habitat, and natural resources), but also indirect impacts. A recent study has shown that premature deaths related to fire emissions of particulate matter (2.5 µm) could double from 10,000s by the end of the 21st century (Ford et al., 2018).

Because of the complex role that fire plays and the impacts it has on society, management strategies consider both before and after fires as well as during active fire. Proactive fire management includes prescribed burning, spreading straw to prevent post-fire soil erosion, felling dying trees, etc.. Reactive fire management include fire suppression tactics such as trenching, back fires, etc..

For both proactive and reactive fire management, managers rely on in situ observations, remote observations via airborne and satellite assets (Barrett and Kasischke, 2013; Elliot et al., 2016; Finco et al., 2012; Oliva and Schroeder, 2015; Parks, 2014; Randerson et al., 2012; Schroeder et al., 2014, 2008), models and meteorological forecasts (Abatzoglou, 2013, 2011). Satellite observations include active fire detections from both MODIS and VIIRS as well as high-resolution airborne thermal infrared imagery on big fires (flown at night) via the National Infrared Operations (NIROPS) program. Current tools include models parameterized from in situ and remote observations. For example, fire behavior models ingest fuel classification maps and meteorological data and to determine how fire will behave (e.g., BEHAVE-plus; Burgan and Rothermel 1984; Andrews 1986, 2009; Andrews *et al.* 2003).

Although there are tools available that provide decision support and enable research, observations of active fire have been limited to extremely fine scales (in situ) or coarse scales (e.g., from satellite or manned aircraft). Neither of these scales are sufficient to resolve process-based understanding of fire that is necessary to advance understanding of the role of fire in the Earth System (Dennison et al., 2017; Soja et al., 2017; Stavros et al., 2017; Sullivan et al., 2017). Thus, intermediate-scale observations from minutes to days and meters to hectares are required, which UAS can enable.

# 2 Relevance to JPL

Developing UAS technologies for fire science and management applications directly relates to the JPL 2025 Quest 7 to explore potential ways that JPL unique expertise can serve the nation and the people. Wildfires are a growing concern that are affecting more people either directly through loss of life and infrastructure or indirectly through degraded air and water quality. Furthermore, the increase in extreme events, that is expected to continue in the future (Barbero et al., 2014; Stavros et al., 2014a), has already cost billions (Figure 2). Thus, developing UAS technologies for fire science and management applications, will continue JPL tradition by tackling issues of national priority and providing help in the civil, commercial, and security sectors. The recently passed congressional bill (H.R. 2862, the Wildfire Disaster Funding Act) demonstrates the national priority of wildfire management.

UAS technologies can also indirectly benefit quests 2 and 4 related to interstellar planetary exploration. One of the biggest challenges with many of the latest technologies developed for artificial intelligence that will be essential to protecting our space assets (e.g., rovers, drones, and satellites) that are far away and require near-real time input, is training data limitation. Moreover, UAS technologies for wildfire recognition, segmentation, and situational awareness directly support JPL's strategic vision of autonomy. Autonomy with learning and adaptation capabilities is crucial for future robotic exploration. By developing these technologies for extreme environments on Earth, we can advance technological readiness such that it could be used for flight eventually.

Not only does development of UAS technologies for fire science and management applications directly relate to JPL quests, it also addresses the need to explore cost-benefit analyses using drone technologies as a part of remote sensing strategies as recommended by the 2017 Earth Science Decadal Survey (p. 6-26; ESAS 2017). Although the Decadal Survey discusses the use of drone technologies specific to water quality, many of the functionalities needed for such technologies may also apply in the wildfire landscape (e.g., mapping in mountainous, forested terrain). Thus, exploring drone technologies for fire science and management, indirectly contributes to the next generation of technologies for addressing broader UAS applications recommended by the Decadal Survey. Additionally, preparedness and mitigation of fire is directly mentioned as a key observation for address key science Question H-4: "How does the water cycle interact with other Earth System processes to change the predictability and impacts of hazardous events and hazard-chains… and how do we improve preparedness and mitigation of water- related extreme events?" and hypothesis H4-d (Important): "Understand linkages between anthropogenic modification of the land, including fire suppression, land use, and urbanization on frequency of and response to hazards."

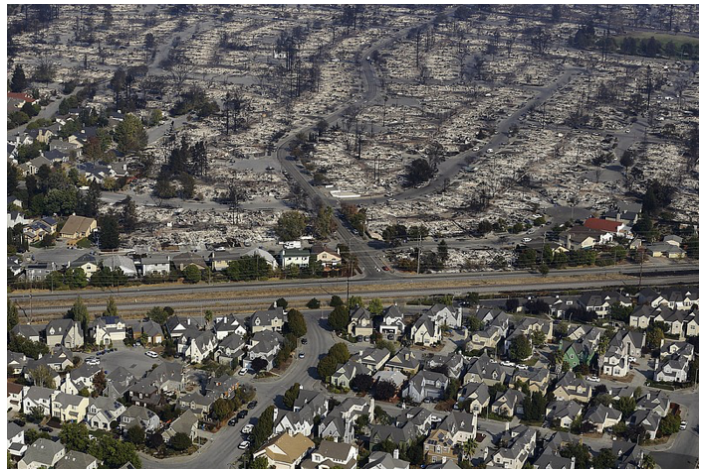

Figure 2 Tubbs Fire, Santa Rosa, California (October 2017) resulted in 149 km$^2$ burned, $\geq$22 people killed, > 5,643 structures incinerated (2,800 ~ 5% of the city's housing), and ~$1.2 billion USD lost. Photo courtesy of kpbs.org.

Lastly, exploring applications of UAS for fire science and management builds on previous work: AI drone technologies for exploring subterranean caves (8x and DARPA), Mars Rover-Copter coordination and navigation, AUDREY AI for first responders (3x and DHS), AI vs. man (Google), Barn Owl (Angel Investor) night flying, TEFIM, and UAVs for agriculture. In addition, this work complements CalTech Center for Autonomous Systems and Technology (CAST) development of emergency response UAS technologies, thus further strengthening the JPL-CalTech partnership.

# 3 Current Applications of UAS for Fire Management

Table 1: UAS classifications for fire science and management applications where types 1 and 2 are generally operated by contractors and provide strategic situational awareness, operate above all other incident aircrafts, and maintain communications with the UAS crew on the assigned Victor (AM) or air to ground (FM) frequencies. Types 3 and 4 are generally agency operated and provide tactical situation awareness along the fireline at relatively low levels (~200'). Types 3 and 4 maintain communications with the UAS crew and are assigned air to ground (FM) frequencies and do not carry transponders or AFF equipment. Source: National Wildfire Coordinating Group (NWCG, 2017)

| Type | Configuration | Endurance | Data Collection Altitude (agl) | Equipped Weight (lbs.) | Typical Sensors |
|---|---|---|---|---|---|
| 1 | Fixed Wing<br>Rotorcraft | 6-24 hrs.<br>NA | 3,000-5,000'<br>NA | >55<br>NA | EO/IR/Multi-Spectral, Lidar* |
| 2 | Fixed Wing<br>Rotorcraft | 1-6 hrs.<br>20-60 min. | 1,200-3,000'<br>400-1,200' | 15-55<br>15-55 | EO/IR/Multi-Spectral, Lidar |
| 3 | Fixed Wing<br>Rotorcraft | 20-60 min.<br>20-60 min | 400-1200'<br><400' | 5-14<br>5-14 | EO/IR Video and Stills |
| 4 | Fixed Wing<br>Rotorcraft | Up to 30 min.<br>Up to 20 min. | 400-1200'<br><400' | <5<br><5 | EO/IR Video and Stills |
| *Contracted aircraft sensors will be determined by the contract specifications. | | | | | |

US Federal land agencies are already using UAS technologies for fire management (Table 1). The current state and advancements are discussed monthly by the interagency UAS Advisory Group (UASAG). The use of drones is very important for providing situational awareness that can be difficult to assess especially when piloted aircraft systems are grounded due to smoky conditions and fear for pilot safety. At least 16 different UAS technologies were used in 2017 including the AeroVironment RavenRQ-11A, 3DR Solo (385), Firefly Y6S (15), Honeywell T-Haw, MLB Super Bat, and the Pulse Vapor 55(1). The most common UAS mass is ≤55 lbs. and costs less than $15k. These technologies are equipped with off-the-shelf sensors such as point & shoot MILC and DSLR cameras (e.g., Sony a6000, Sony A7r, and Sony RXIRII), multispectral sensors (e.g., Slantrange 3P, MicaSense RedEdge, Micasense RedEdge M, and Parrot Sequoia), high-definition video (e.g., Gimbaled GoPro Hero 4), and thermal sensors (e.g., Gimbaled FLIR Vue Pro R and Colibri). Current partners for providing these technologies for fire management include DJI, DOD, 3DR, Bremor, Lockhead Martin, Boeing, and Precision Integrated.

The current applications of UAS are primarily for active fire management and rely on pilots. In particular, the information obtained during active fire of greatest interest are the location of spot fires, fire front location, flame lengths, particulate matter and CO concentrations, vertical surface assessment (steepness of terrain and connectivity of fuels), as well as proximity of the fire to critical infrastructure and iconic historical sites. Unfortunately, the current state for providing this information from UAS technologies requires pilots and data managers to spend about 2 hours to process the data for every 15 minutes of flight. As such, the UASAG is looking for smart solutions to store and process the data into meaningful information for real-time use. Lastly, there is a sociocultural resistance to use of drones as there is much concern of privacy of information and methods to ensure protections of that privacy. Furthermore, non-participating (i.e., not part of the fire management agency working a fire) drones are restricted and it is difficult to justify the use of drones by the fire management agency without having robust systems integrated for identification and communication with the manned aircrafts used on the fire.

Some limitations of the current UAS practices are: the maintenance (training and UAS parts) and operating costs (work effort) of the UAS and pilots, the FAA restriction to maintain line of site, data processing and dissemination, and the sociocultural acceptance for using the technology. Fortunately, policy hurdles are becoming easier to manage as the FAA expands their applications. Specifically, NASA Ames has worked with FAA to reduce the need for prior approval when flying less than 400 ft altitude for emergency management situations.

Areas of future research being explored are how to extend flight duration and aircraft endurance, working offline from the DJI network, an affordable small, multi-rotor aircraft, equipping UAS with ADSB/mod-C for traffic collision control, and use of UAS for tactical missions like search and rescue missions and cargo/water dropping and other aerial applications.

# 4 Future Applications of UAS for Fire Management

There are a number of applications in fire management for which UAS could prove valuable, specifically: pre-fire fire danger assessment, active fire situational awareness, active fire logistics transport, active fire suppression, and post-fire damage assessment. Each application was discussed with the federal Interagency UAS Advisory Group (UASAG) for fire operations. Based on the current tools that are available and how operations are conducted, the UASAG provided a qualitative response as to the utility of each application. Here we provide a brief description of the application, the current technologies used by fire operations, and report the UASAG utility rating as low, medium or high. The needed functionalities of UAS for each application are reported in Table 2 and discussed in detail in Section 6.

## 4.1 Pre-Fire: Fire Danger Assessment

There are three characteristics that result in sustained combustion: fuels, heat, and oxygen; once combustion occurs it can result in a wildfire, which is driven by fuels, topography and weather (Agee, 1993). Since fuels play an important role in both combustion and driving wildfire behavior, understanding their current condition, structure, and types is essential for predicting both how at risk an area is for sustaining fire, and in how that fire will behave. At present, US wildland and fire management agencies use the National Fire Danger Rating System (NFDRS; Bradshaw et al., 1983) to predict fire danger out to 1-10 days in advance. NFDRS is based on empirical relationships (Rothermel, 1972) and relies on input values from the USGS EROS LANDFIRE fuels database (Nelson et al., 2014; Rollins et al., 2006; Rollins, 2009), parameterizations of fuel classes based on Remote Automated Weather Station (RAWS) data and field data, and national-scale reanalysis data (Abatzoglou, 2013, 2011; Holden and Jolly, 2011). These predictions often have spatial resolutions on the order of square kilometers. This resolution is enough to release public advisories of fire danger (often listed at the entrances to federal land).

More frequent and higher spatial resolution models (e.g. BEHAVE-plus - Andrews, 2009; CAWFE - Coen, 2013) are used to predict active fire behavior and inform fire operations managers. As such, the UASAG ranked this application as providing low utility on top of current operations.

## 4.2 Active Fire: Situational Awareness

During active fire, there is a pressing need to understand where the fire is, where management troops are, and how they relate to each other on the landscape. At present, active fire situational awareness comes from satellite data of active fire (e.g., VIIRS and MODIS), weather data, topographic maps, and radio inputs from fire management troops on the ground, in aircrafts (e.g., dispensing fire retardant or the National Infrared Operations - NIROPS), and from piloted UAS equipped with cameras. A fire operations manager is expected to aggregate this information and make decisions about resource allocation. To do this, he may also rely on inputs from fire behavior models (e.g. BEHAVE-plus - Andrews, 2009; CAWFE - Coen, 2013). Unfortunately, there is no single interface to aggregate this information beyond the capacity of the leading operations manager.

Table 2: In order for UAS to be used for any one application, there are a number of functionalities (i.e., use cases) needed. Section 6 describes each use case in detail, here we provide a link between each application and the use cases.

| Application (utility) | Autonomy | Distributed Autonomy | Autonomous Sensing | Swarm Mechanics | Data and Information Systems | Hydro Propulsion | Mobile, AI-deployed Infrasound Projectors |
|---|---|---|---|---|---|---|---|
| Pre-Fire: Fire Danger Assessment (low) | X | X | X | X | X | | |
| Active Fire: Situational Awareness (high) | X | X | X | X | X | | |
| Active Fire: Logistics Transport (high) | X | X | X | X | | | |
| Active Fire: Suppression (low) | X | X | X | X | X | X | X |
| Post-Fire: Damage Assessment (high) | X | X | X | X | X | | |

© 2018, California Institute of Technology. Government sponsorship acknowledged.　　　　　Page | 6

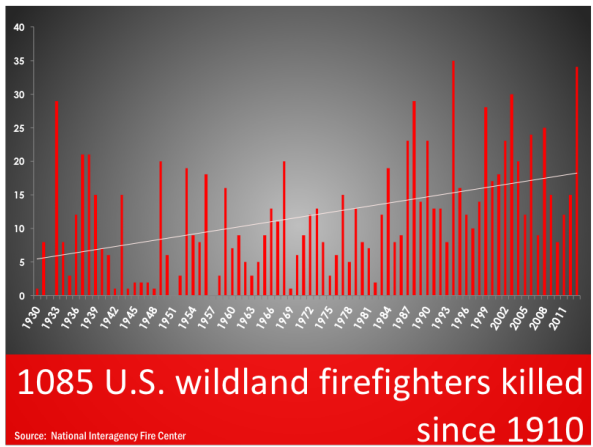

Figure 3: There is an increasing trend in wildland firefighter mortality since 1930. Source: National Interagency Fire Center.

When situational awareness was discussed with UASAG, four specific applications were discussed: 1) escape route mapping, 2) search and rescue - person identification (no GPS), 3) person geolocation (with GPS), and 4) local environment mapping. First and foremost, the UASAG stressed that the highest priority was protecting human life (Figure 3). The first three applications all emphasize this point, but geolocation of personnel garnered the most interest as knowing where resource assets are in relation to the moving fire is an area of great unknown. The UASAG also showed great interest for improving localized environmental awareness (e.g., geolocation of fire front, how hot it is, how fast and in what directions the winds are moving, etc.). The emphasis of this was on the need for high-powered compute to transform data into information that could be used in real time. Recently, JPL's Assistant for Understanding Data through Reasoning, Extraction and Synthesis (AUDREY) AI, funded by DHS, is being developed for situational awareness for first responders by ingesting unstructured data in real time and producing insight to support decision making of first responders (McKinzie 2018). The data that is currently collected takes too long to process and is thus most useful based on the real-time human interpretations through live UAS feeds. Not only is the challenge with UAS in the data and information systems, but also in the upfront costs to train and limited availability of UAS pilots for fire management operations. As such, affordable AI drone technology that removed these barriers would enable use of UAS to improve situational awareness on more fires, and not merely the largest and most extreme.

## 4.3 Active Fire: Logistics Transport

When fighting a fire, there are a two major costs: 1) resources (materials, labor, etc.), and 2)

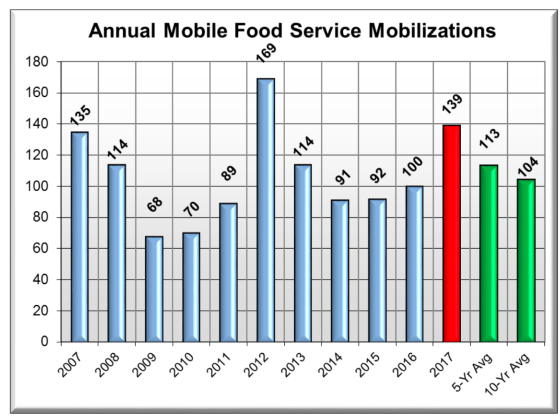

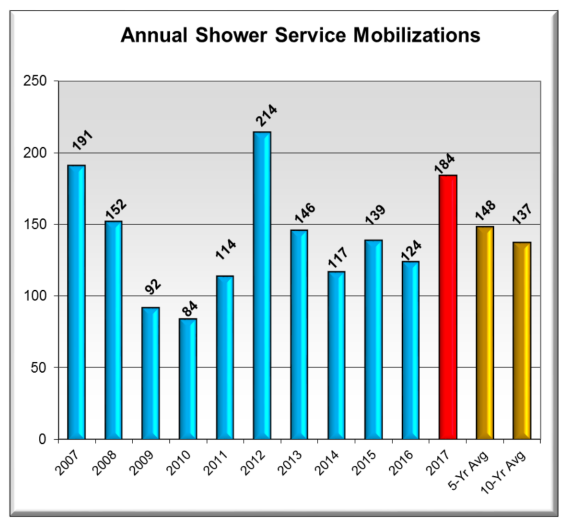

Figure 4: UASAG reported annual mobilizations of resources around active fires by year. These tables are copied from a National Interagency Fire Center (NIFC) Predictive Services report: https://www.predictiveservices.nifc.gov/intelligence/2017_statssumm/resource_charts_tables17.pdf

 

moving resources around. Broadly, resources can include aircrafts, personnel, radios, firefighting equipment, food, mobile showers, etc. There was discussion by UASAG of the ability of AI-piloted UAS to help move resources around. This could substantially help re-allocate man-hours and transport resources to help with suppression activities (Figure 4). To help with this activity, a UAS would not only require AI technology for driving the drone, but also enough power to deliver the needed resources. At present, resources are delivered via ground or air transport services, both of which require access and manpower. As such, the UASAG has ranked this is an area of medium priority.

## 4.4 Active Fire: Suppression

Fire suppression is the most intuitive response to wildland fire. In 1910, after the Great Fire in Northern Idaho and Western Montana (~ 3,000,000 ac ~ 12,100 km$^2$), the United States launched the 10 AM fire exclusion policy - all fires must be out by 10 am the next day. By the 1970s, we learned that fire was essential to maintaining the health of many ecosystems (Mutch, 1970), as such the national policy shifted to allow fires in remote locations to burn. Even with this policy, however, there are still fires that we must suppress. The challenge is that wildfires can spread very quickly (e.g., 3 miles/hour). If it takes approximately 15 minutes for the average wildland fire to be reported in the United States, and at least 45 minutes for fire management resources to be deployed to the remote location of the fire, the fire will have grown. For a back of the envelope calculation, we could assume that a fire grows half as wide as it does long, which would result in a fire perimeter of 9.7 miles. Even with some of the best resources in the world for suppression averaging production line rates of 100s ft/hr (~ 10s m/hr; Table 2), a fire growing that fast cannot be suppressed. In these cases, fire operations managers will decide where assets are that need protecting, where evacuations need to occur, and the most strategic places to deploy resources for protecting assets. Because of this, the UASAG decided that fire suppression technologies would provide lower utility than the other applications discussed.

## 4.5 Post-Fire: Damage Assessment

Wildland fires can cause damage long after they have been contained, as such all federal lands are expected to provide an environmental impact report of each fire within 7 days of 100% containment. Environmental impacts can include soil damage, potential for erosion and debris flows, hazardous trees, landslide potential, water quality impacts, and loss of endangered species habitat. These reports are conducted by the Burn Area Emergency Response (BAER) Teams and rely on field data collections, remote sensing datasets (e.g., Sentinel and Landsat), and models (e.g., the NASA Applied Science Wildfire Program funded Rapid Response Erosion Database - http://rred.mtri.org/rred/).

The remaining challenge is in personnel safety to cover the fire extent immediately post-fire. The post-fire landscape can be quite dangerous as often there are still smoldering and burning hot spots within the perimeter, hallowed ground from the burning of tree roots that can result in sinkholes, unstable terrain, and hazardous trees. AI piloted drones that could quickly image and process to value-added information these landscapes, would have the potential to add security and expedite post-fire damage assessment. As such the UASAG qualified this application as providing high utility.

 

Table 3: Sustained Line Production Rates are the expected length of fireline, or fuel break created by clearing the land, that can be constructed by a crew in a given time. The following table is from https://www.nwcg.gov/sites/default/files/publications/pms210.pdf - pg. 120.

| Sustained Line Production Rates of 20-Person Crews in Feet per Hour* | | | | |
|---|---|---|---|---|
| **Fire Behavior Fuel Model** | **Type I Direct** | **Type I Indirect** | **Type II & II IA Direct** | **Type II & II IA Indirect** |
| 1 Short Grass<br>2 Open Timber Grass | **1,122**<br>(792-1,386) ** | **627**<br>(508-746) | **627**<br>(174-660) | **285**<br>(174-380) |
| 4 Chaparral | **436**<br>(330-528) | **330**<br>(178-482) | **449**<br>(80-640) | **272**<br>(178-376) |
| 5 Brush | **1,089**<br>(924-1,254) | **323**<br>(244-403) | **471**<br>(30-682) | **277**<br>(178-376) |
| 6 Dormant Brush<br>Hardwood Slash | **1,089**<br>(924-1,254) | **323**<br>(244-403) | **471**<br>(30-682) | **277**<br>(178-376) |
| 8 Closed Timber Litter<br>9 Hardwood Litter<br>10 Timber Litter & Understory | **693**<br>(594-792) | **455**<br>(396-515) | **447**<br>(370-448) | **378**<br>(255-452) |

* Based on San Dimas Technology & Development Center, Tech Tip – 1151-1805P, Fireline Production Rates, 2011. No data was collected in fuel models 3, 7, and 11 – 13.
** Numbers in parentheses are expected ranges of line production.
IA = Initial Attack

# 5 Use Cases and Needed Functionality: Current State and Advancement

Each application has technology use cases, which we articulate here. We describe the use case and the needed functionalities as well as the current state and advancements needed in order to actually use the technology for each of the aforementioned applications. For all software and hardware, we report the technology readiness level (TRL), and for information systems development (i.e., the transformation of raw data to relevant, value-added information), we report the application readiness level (ARL). TRL and ARL are similar, but TRL is used to report the readiness of technology for for flight, while ARL is used to report the readiness for a data and information system for operational use in decision making.

## 5.1 Autonomy

Autonomy is the ability of a robotic agent to operate and make decisions without direct human intervention and control from outside. To provide Autonomy three functionalities need to be developed: 1) navigation localization, 2) distributed autonomy, and 3) autonomous multi-sensing technologies.

### 5.1.1 Scalable, robust localization and mapping

A Micro Aerial Vehicle (MAV), a small, lightweight UAS, operating in a remote environment, needs to obtain its location with high accuracy and navigate toward a goal target only by taking aerial images of the environment and registering them to a provided aerial map of the environment. Lack of salient features and landmarks, e.g., during poor visibility due to dense smoke and high contrast in images of fire, makes the tasks of vision-based localization and navigation very challenging. Vision-based localization and mapping in dense smoke and poor visibility by relying on vision and thermal imaging is an ongoing research. At present, simultaneous Localization and Mapping (SLAM) algorithms are not robust and require large-heavy computing systems. As such, this functionality is currently at TRL 3. With low seed investment, this technology could be adapted to develop outlier-tolerant vision-based graph optimization, develop optimized keyframe selection, develop scalable graph optimization methods, thus advancing to TRL 6.

### 5.1.2 Scalable Autonomous Motion Planning and Obstacle Avoidance

Motion planning in an unstructured environment is the capability that enables a group of autonomous robotic agents (MAVs) to collaborate autonomously in an unknown and unstructured environment without crashing into each other or into obstacles. It also enables the MAVs to plan the most efficient path from point A to point B which translates into higher operational speeds and lower costs. At present, slow and unreliable collision avoidance has been proven at hundreds of meter range in a laboratory setting or a controlled environment. As such, this functionality is currently at TRL 4. With low seed investment, tested fast-motion and path planning can be adapted for and tested in unstructured, cluttered environments (i.e., forests, wildfires, etc.), thus advancing to TRL 6.

### 5.1.3 Autonomous Multi-Sensing technology

Autonomous multi-sensing technology is the capability of autonomously detecting regions of interest and collecting critical information in an unknown environment using active and passive sensing technologies specifically integratd in the onboard data systemt to inform the motions/actions of the MAV (i.e, different from Section 5.3.1). An example of passive sensing technology is a MAV equipped with RGB or thermal cameras and can autonomously detect hot spots in an environment and change its direction or direct its onboard cameras toward the detected hot spot to gather more detailed and high-resolution information. An active sensing is a radar, or LiDAR, that can be used to map the environment as the MAV flies overhead. MAV equipped with fire monitoring sensors (visual, thermal, infrared, anemometer, etc.) can be deployed to scan an environment and search for signs of anomalies (high heat signature, smoke, high contrast in visual imaging, etc.). Once an area with an anomaly has been flagged, more MAV can be dispatched to the location to study and collect detailed information. Before this can happen, all sensing technologies must communicate the environment status to inform the MAV where to fly. This level of autonomy and collaborative monitoring/surveying of a remote and unstructured environment has not been done in a real forest setting. The current state for perception can be deployed indoors, outdoors and in low visibility environment by relying on sensing technologies in the visible to the thermal infrared, LiDAR and radar. The perception performance for object detection and segmentation has greatly improved by implementing deep learning-based computer vision algorithms, including conditional generative adversarial networks (CGAN) and Mask RCNN (Yun et al. 2018). However, the reliability of perception algorithms for survey fire in a forested environment, has not really been explored. As such, this functionality is currently at TRL 2. With low seed investment, multi-sensor fusion (e.g., using LOAM, MSCKF, etc.) in combination with advanced deep learning algorithms is expected to achieve increased reliability in forested, perceptually-degraded environments, thus advancing to TRL 6.

## 5.2 Distributed Autonomy

In a large forested environment, it is critical to survey and map the entire environment in the shortest amount of time. This can be achieved by deploying multiple MAVs in the environment and tasking each with mapping a section of the forest. The data collected by all the MAVs needs to be stitched together so a larger map can be obtained. There are three functionalities needed for the distributed autonomy use case: 1) multi-MAV mission/motion planning, that is the ability of MAVs to work together without crashing into each other and without having too much overlap between the environment they are mapping. 2) distributed and collaborative SLAM on MAV swarm, that is the ability to collaboratively create a large map of the environment, and 3) resilient communication network on MAV swarm, that is the capability to communicate between MAVs and the control center to rely sensitive data and information.

### 5.2.1 Multi-MAV Mission/Motion Planning

Motion planning for a Multi-MAV includes avoiding collisions with each other and with other obstacles in difficult conditions like dense fog and poor visibility, and must be able to efficiently distribute the tasks among themselves so the task can be accomplished in the most efficient way. The current state of multi-MAV mission/motion planning has been limited to an academic lab setting in simple scenarios (e.g., unobstructed and clear visibility). As such, this functionality is currently at TRL 3. With low seed investment, robust planning and seamless interaction with human operators could be achieved by researching and developing scalable and real-time MAV swarm motion, developing multi-robot collision avoidance, and bridge the gap between high-level operator commands and low-level motions and configuration of MAVs in perceptually-degraded environments, thus advancing to TRL 6.

### 5.2.2 Distributed and Collaborative Simultaneous Localization and Mapping (SLAM)

Wildland fire monitoring using a swarm of MAV could provide flexibility, robustness, modularity, and scalability. Specifically, small MAV can organize into networks composed of as many nodes as necessary, enabling system extensibility. The cooperation between multiple MAV, however, depends on a distributed autonomy MAV scheme. For example, in a forest-fire detection, each MAV is equipped with visual and thermal sensors to detect a fire. Once a fire is detected, an alarm along with the geotagged location of fire is transmitted to the Control Station. The alarm is used to dynamically command other MAV equipped with specific fire sensors, which are sent to the location of the potential alarms for confirmation and more detailed information about the size and direction of the fire. The data gathered by a group of MAV working collaboratively, is combined using fusion techniques in order to reliably confirm or discard the fire alarm and create a high spatial and temporal resolution of the environment. The constructed spatio-temporal map can be further utilized for real-time simulation of fire propagation and resource allocation. Therefore, firefighters will have better situational awareness and can even predict situations.

Since ego-motion estimation and map-building are key in enabling autonomous navigation, SLAM is fundamental in Robotics with most of the literature focusing on SLAM from a single MAV. As such, the current state of such distributed autonomy does not exist beyond structured and controlled environments (i.e., building a map of a structured outdoor environment, an office or small building, hallway, etc.) (TRL 4). With low seed investment, robust distributed mapping in unstructured environments could be developed and tested, thus advancing to TRL 6.

### 5.2.3 Resilient Communication Network on MAV Swarm

For a group of MAV to collaborate efficiently and safely, there needs to be a stable and reliable communication link between all robotic agents and the control center, so that at any point in time they can communicate critical information (i.e., their location, observations from forest, fire status, etc.). The current state for such technology includes JPL's Disruption Tolerant Network (DTN), which provides a general-purpose network/transport-layer service that is logically similar to what Transmission Control Protocol/Internet Protocol (TCP/IP) provides for the terrestrial Internet, but suitable for use in the space environment. In addition to the basic store-and-forward internetworking service, DTN also provides efficient reliability, security, in-order delivery, duplicate suppression, class of service (prioritization), remote management, and a 'DVR-like' streaming service, rate buffering, and data accounting. DTN leverages asymmetric and time-disjoint paths. As such, this functionality is currently at TRL 2 for a remote, cluttered forest environment where wildland fires occur, but with low seed investment, the JPL DTN could be tested in the appropriate environment, thus advancing to TRL 6.

## 5.3 Autonomous Sensing Technology

In order for drones to not only navigate and map autonomously, they must collect sensory inputs from the environment around them. As such, there are two functionalities needed from sensing technologies for drone automation: 1) fire hot spot detection, and 2) wind speed and direction.

### 5.3.1 Fire Hot Spot Detection

Hot spot detection provides valuable input for motion planning and navigation of any single MAV and particularly for distributed swarms as it informs each MAV where to focus efforts. However, artificial intelligence must be applied to improve the algorithms for detecting hot spots from sensory data. Furthermore, hot spot detection would be invaluable to relay back at the command center to inform where exactly the fire is moving across the landscape. The current state by fire management using MAV relies on thermal infrared cameras, but does not quantify or convert imagery into information (e.g., perimeter mapping, intensity, geolocation, etc.) for real-time decision making. Specifically, the decision process relies on the expert judgement of professional drone pilots. The current state of using thermal infrared imagery for autonomous path planning and navigation of UAS is an ongoing research problem (TRL 2). With low seed investment, thermal infrared imaging could be integrated and tested for a perception-poor environment with hot spots, thus advancing to TRL 6.

### 5.3.2 Wind Speed and Direction

Measuring wind speed and direction is extremely important for autonomy because it informs by how much to correct when offset by changes in wind. In a wildfire situation, where fire creates its own microclimate with strong winds (Coen et al., 2018), using on wind information can provide input to the controller that is responsible for balancing MAV in the air. At present, fire management relies on pilots to correct for winds in a turbulent fire environment. Anemometers exist, but have not been integrated with off-the shelf UAS control systems. As such, the current state for anemometer integration with the controller to enable autonomous and robust navigation in wildfire microclimates is TRL 2. With low seed investment, anemometers could be integrated and tested in a turbulent fire environment, thus advancing to TRL 6.

## 5.4 Swarm mechanics

A swarm of drones engaging an active fire would be subject to several effects that can potentially be very demanding, as the stability and safety of each vehicle is affected. An active fire generates its own microclimate (Coen et al., 2018), with strong convective currents generated by steep temperature and pressure gradients near the surface, as well as at high altitude. Also, visibility is impaired due to the dense smoke, requiring imaging stabilization techniques both in the visible and infrared bands. Stable hovering and control in high and variable wind thus becomes a priority, driving the propulsive needs of the vehicle, and demanding high situational awareness for real-time on-demand, changes of trajectory. Consequently, there are four key functionalities needed for enabling swarm mechanics (independent of the autonomy needed - Section 6.2.ii): 1) environmental awareness while hovering in high and variable wind, and 2) propulsion for attitude control and hovering in high and variable wind.

### 5.4.1 Environmental Awareness

Environmental awareness while hovering in a high and variable wind environment is needed for swarm mechanics in order to inform propulsion and attitude control. The current state is for guidance, navigation and control in low wind environments with GPS-assisted algorithms that rely on visible/infrared cameras, as such this is TRL 2. With low seed investment, coordination algorithms could be developed for handling multiple assets in high adverse winds and work in real-time in a GPS-denied environment, thus advancing to TRL 4.

### 5.4.2 Propulsion and Attitude Control for Hovering

Environmental awareness while hovering in a high and variable wind environment is needed for swarm mechanics in order to inform propulsion and attitude control. The current state is for guidance, navigation and control in low wind environments with GPS-assisted algorithms that rely on visible/infrared cameras, as such this is TRL 2. With low seed investment, coordination algorithms could be developed for handling multiple assets in high adverse winds and work in real-time in a GPS-denied environment, thus advancing to TRL 4.

## 5.5 Data and Information Systems

Data can be collected in many forms, but it is often useless in its raw form. The value of data can be found after processing it into information specific to the decision-making activity at hand. For example, on fires, thermal infrared imaging is taken on drones today, but this merely provides visual guidance, and pilots/operations managers are left to fill in the blank on where a thermal anomaly is geographically occurring in relation to resources or in relation to the fire. Information can be derived as patterns found in data, but the most useful information is derived from understanding what decisions are made, what knowledge was used to inform them, and the consequently what information would elucidate how to apply that knowledge to make a decision. At present remote sensing data are available for fire management, however much of it is not in a form that can be used in real-time operational decision making as a fire moves across the landscape at a speed of kilometers per hour. Thus, data and information system functions needed for (before, during and after) fire management, include: 1) Vegetation Stress, 2) Escape Route Mapping, 3) Person Identification (no GPS), 4) Person Geolocation (with GPS), 5) Local Environment Mapping, and 6) Damage Assessment Mapping.

### 5.5.1 Vegetation Stress

The heat signature of plants can inform how stressed and dry they are, and consequently inform areas of highest fire danger in a pre-fire environment. Drones with a thermal infrared sensor can provide valuable data that can then be transformed into information of plant stress (e.g., ECOSTRESS Level 4 PT-JPL ESI ATBD, D-94647), which can inform areas that are more susceptible to burn than others. The current state of application for this technology uses multi-band thermal infrared sensors and has been demonstrated on drones in agricultural environments (ARL 3). With low seed investment, such an algorithm could be adapted for a forested environment, thus advancing to ARL 6.

### 5.5.2 Escape Route Mapping

In this context, escape route mapping applies to the application of personnel working on the fire needing a path to escape when fire moves into their area. The current state relies on hand-held devices that can have the park map and routes loaded. Unfortunately, this does not include information on the fire (where it is, areas that may be more dangerous after the fire has passed, etc.). Furthermore, fire perimeter mapping relies on either course scale satellite overpasses active fire detections that merely detect the flaming front (i.e., products do not update fire progression) or night-time airborne perimeter maps that are only generated for the most severe fires, as such this functionality of information systems is ARL 1. With low seed investment, process knowledge

 

mapping to identify needed information, and algorithm development of that information from data could be validated, thus advancing to ARL 4.

### 5.5.3 Person Identification (no GPS)

The operational fire management environment is complex with many moving parts. Often one of the most challenging tasks is knowing where all assets are in relation to one another and in relation to roads/access routes and the fire itself. Search and rescue tracking would rely on thermal imaging to identify and distinguish missing persons. The current state of search and rescue of civilians (e.g., campers, hikers, etc.) during fire is challenging and does not have clear solutions when mobile phone reception is poor because often evacuation alerts are pushed to their phones. Image classification technology has been developed and tested on UAVs (Rudol et al., 2008), as such this application is ARL 3. With low seed investment, such algorithms could be adapted for smoke-obscured, forested landscapes and tested for deployment in fire management information systems, thus advancing to ARL 6.

### 5.5.4 Person Geolocation (with GPS)

For the active fire application, person geolocation refers to personnel position in the context of the all other assets and on the landscape. The current state is for all fire management personnel carry a GPS, but this data does not necessarily read into a single interface that collocates all other relevant information (e.g., other assets, topography, infrastructure, and the fire). It is thus the job of the operations manager to aggregate this information from different sources. JPL recently deployed the beta version of the AUDREY dashboard that integrates and visualizes relevant information for Grant County's dispatcher center, funded by DHS Next Generation First Responder (NGFR) program. Some commercial companies have begun to develop systems that aggregate relevant information into a single map interface (e.g., Tecnosylva), as such this application is ARL 3. With low seed investment, personnel geolocation could be integrated with other relevant information and tested in the decision-making context, thus advancing to ARL 7.

### 5.5.5 Active Fire Local Environment Mapping

Fusing multiple sensor data (anemometer, heat signature, size of the fire, etc.) can provide valuable information about fire intensity, front location, hot spot detection, spot fires, and rate and direction of spread. This specific use case integrates use case 5.3.1 Fire Hot Spot Detection with an user interface to provide information relevant for fire management decision support. The current state relies on expert knowledge of how fires move through a landscape, topographic maps, meteorological data derived from weather stations often kilometers away from the fire, satellite active fire detections, and aerial imagery from piloted drones. The current state uses all of this data and information, but relies heavily on an operations manager to collocate and synthesize the information on the fly without being integrated into a single information system, as such this information system is ARL 1. With low seed investment, this information could be integrated into a single information system, process knowledge mapping, and algorithms for transforming drone sensing technology data into value-added information developed and validated, thus advancing to ARL 4.

### 5.5.6 Damage Assessment Mapping

Within 7 days after a fire is contained, US federal lands must provide a damage assessment. This assessment includes vegetation burn severity and soil burn severity mapping as well as concerns about hazardous trees, potential debris flows, damage to water reserves, and erosion. The current state relies on the Landsat constellation (including Sentinel 2) and models to generate maps as well as field observations. These environments can be particularly dangerous for field crews as there are often hotspots that are still burning, trees may fall over, or the ground can collapse as root systems have burned leaving unsupported tunnels. The use of a drone to survey post-fire environments could prove particularly valuable, however the current state does not process data at a rate fast enough to convert the raw data into information relevant for the final report (ARL 1). With low seed investment, damage assessment experts could be interviewed, process knowledge mapping conducted to define information product specifications, and algorithms could be tested for providing value-added information, thus advancing to ARL 4.

 

## 5.6 Hydro Propulsion

A firefighting system utilizing pressurized water both as propellant for mobility and as a dousing agent to extinguish fires could be an effective tool for firefighting. Commercial products such as the Jet-Lev backpack demonstrate the feasibility of aerial mobility of a robotic drone performing water disbursement. Analysis has shown such water 'drones' to be feasible with effective ranges of 100 meters or more. The water drone platform would be inertially stable as it would mimic the physical architecture and control aspects of quadrotors. This system would allow the water drone access to areas dangerous for human firefighters; brush fires on steep slopes, tall buildings fires, chemical fires, marine fires, and fires in radiation environments. These systems can be anchored on trucks providing mobility to the fire scene and connect to existing fire hydrants. Pumps would further pressurize the water to the necessary propulsive force. On board sensors would detect infrared and optical signatures of fire locations. These systems could operate both through operator control (teleoperation) or autonomously (Section 6.1) using sensor data (local infrared area maps, GPS). Such craft could have grappling manipulators to anchor themselves in strategic locations while providing water flow. Multiple water drone operations can be coordinated in the same area. Such a hydro-propelled drone would require four functionalities: 1) a hose, 2) water valves, 3) thruster/attitude control and 4) a pressurized water source.

### 5.6.1 Hose

To connect a hose to a UAS for wildfire use would require a flexible, lightweight hose with an integrated electrical tether for control. The current state uses fire hoses composed of canvas and rubber, but has not be integrated with the relevant supporting elements. Thus, despite the high TRL of fire hoses, hoses for this application are at TRL 1. With low seed investment, the hose material could be fabricated and a hose prototype produced and pressure testing applied, thus advancing to TRL 4.

### 5.6.2 Water Valves

For a hydro-propulsion UAS, the water valves would need to be lightweight with quick response to electronic control panel. The current state for gating and control of pressurized water currently exists in industry and is used in the Jet-Lev backpack, as such this is TRL 2. With low seed investment, high-strength, lightweight, fast-response molded valves could be adapted for UAS, thus advancing this technology to TRL 5.

### 5.6.3 Thruster Control/Attitude Control

A hydro-propulsion UAS would require that each of the jets be used for thruster and attitude control. The current state for such technology is quite mature (e.g., drones, quadcopters) and has been used in various industries, however applying it for hydro-propulsion is still in concept, as such this is TRL 2. With low seed investment, existing drone 4-propeller actuator control algorithms could be adapted for water propulsion, thus advancing to TRL 5.

### 5.6.4 Pressurization

In order to fly and maneuver a hydro-propelled UAS, adequate pressurization would be required. The current state of turbopumps to pressurize water exists in industry, but has not been integrated into UAS technologies, as such this is TRL 2. With low seed investment, current pumping systems could be adapted to produced pressurized water (>100 psi) on a UAS, thus advancing to TRL 5.

## 5.7 Mobile, AI-deployed Infrasound Projectors

Infrasound waves are pressure waves below ~20 Hz. They display a low amount of attenuation, and can propagate across long distances, unimpeded by topography. The infrasound wave similar to the P-wave in seismology, with a similar frequency range. Sources of infrasound are open ocean waves, surf, atmospheric nuclear tests, earthquakes, avalanches, meteors, tornadoes, auroras, jets, and volcanoes. They are not restricted by clouds, but affected by wind and temperature gradients. The idea of fire being affected by sound was discovered by John Tyndall in 1874 (Tyndall, 1874). The phenomenon of sound interacting with flames exists in the study of combustion instabilities in aircraft engines and rocket propellant (Humphrey et al., 2016), flame manipulation (Baillot and Demare, 2002) and extinction (Saito et al., 1998). DARPA (DARPA, 2013) showed that it might be possible to control flames by electromagnetic and acoustic waves which will interact with the plasma in a flame, and

 

conducted its "Instant Flame Suppression" program to help extinguish fires in small spaces since 2008 after a shipboard fire on the USS George Washington, which burned for 12 hours and caused an estimated $70 million in damage. Nonchemical flame control using acoustic waves from a subwoofer and a lightweight carbon nanotube thermoacoustic projector has been demonstrated experimentally (Aliev et al., 2017; Park and Robertson, 2009). Laminar flame control and extinction were achieved using a thermoacoustic 'butterfly' projector based on freestanding carbon nanotube sheets. Flame suppression by adiabatic cooling using pure sound waves requires a sound pressure level of at least 158 dB/m. For small laminar flames, the displacement of the flame at low frequencies stretches the flame and reduces its temperature. The low frequency sound waves provide large flame displacement allowing sufficient time to cool the fuel source by supplying colder air. For large-scale fires, short turbulent gusts of sound wave created in narrow tubes can be an efficient acoustic suppression method. A prototype tested at the IMS infrasound array I59US demonstrated the ability to insonify all elements of the array from a standoff distance of 3.8 km. Signal-to-noise ratios of continuous wave signals ranged from 5 to 15 dB, indicating the utility of this source to transmit controllable infrasound signals over distances of 5 km. In order for a mobile, AI-deployed infrasound projector to be useful for wildland fire there are three functionalities that would need to be developed: 1) long-range duration and operation, 2) coordination and directionality, and 3) available acoustic power.

### 5.7.1 Long-Range Duration and Operation

In order to deploy infrasound for controlling wildfires, the instrument would need to be deployed kilometers from the flaming front. The current state applying infrasound for flame control has been demonstrated at a few meters (Ali 2017), as such this is TRL 1. With low seed investment, large-scale demos in open space could be conducted to test the range, power, number of sources and system effectiveness of deploying infrared for flame control from kilometers away, thus advancing to TRL 3.

### 5.7.2 Coordination and Directionality

For this technology to be effective on wildfire, which can have flames 10s to 100s feet tall, there would need to be many projectors (i.e., a swarm), which would need to accurately project infrasound and coordinate their positioning. The current state for swarm control has low accuracy and is thus at TRL 1. With low seed investment, the Guidance, Navigation, and Control algorithms could be tested for precise swarm control, thus enabling maximum coverage and advancing to TRL 3.

### 5.7.3 Available Acoustic Power

In order to deploy projectors for wildfire flame control, the technology would need to have an acoustic power of hundreds to thousands of Watts. The current state for projectors has only demonstrated power of a few Watts (Par 2009), as such this is TRL 1. With low seed investment, the sub-woofers could be developed and tested to provide hundreds to thousands of Watts, thus advancing to TRL 3.

# 6 Recommendations

Ultimately, high return on investment is found at the intersection of impact, maturity and feasibility (Table 4). **Impact** can be broadly defined as the impact of the investment in terms of lives saved, dollars saved, risk avoidance, etc.. Assessing impact can be quite a cumbersome task, but for the purposes of this report, we define it as the impact it would have on fire management operations and the likelihood that the value of the tool would motivate the fire management community to invest in technology transfer. This kind of buy-in from this community is essential because it directly relates to three areas for consideration when anticipating a future business environment: policy, economics and sociocultural (Aguilar, 1967). Ultimately, developing a technology will require the community to create demand for the product, which they can only do if 1) they are interested in and 2) it is affordable within their means (Economics). For the demand to be created, the stakeholder community that trusts the adopted technology to add value to the decision-making context in hand (Sociocultural). Lastly, for this change to occur, there must not be any laws that obstruct the use of that technology (Policy). If there is enough demand for the technology, the laws can change to accommodate them. For example, social media changed the way in which companies gather customer data for sale to third parties, and it was only after there was a demand by the stakeholders that the laws were re-evaluated by the courts. With this in mind, *impact* was assessed based on the qualitative ranking of priority by the fire management community UAS Advisory Group (UASAG). UASAG is an interagency advisory group that is responsible for determining how and when UAS technologies can be used in fire management. We use the standard NASA Technology Readiness Level (TRL; for hardware/software) or Application Readiness Level (ARL; for data information products) to assess the current state of *maturity* for each use case. Lastly, *feasibility* is identified by the change in TRL/ARL given a JPL initial investment ($k) and the number of potential future funding sources. Delta TRL/ARL of 3+ is considered a good investment (green), 2 is considered cautionary (yellow), and 1 is risky (red). JPL investment costs are defined by RTD funding such that a topical RTD at $200k/yr for 3 years ($0- 600k) is considered a good investment (green), an SRTD at $600k/yr for 3 years ($600-$1800k) is a cautionary investment (yellow), and anything more than that is risky (red). If only one future funding source could be identified it was marked as low (red) likelihood of return on investment. If it had at least 2, it was marked as medium (yellow), and if it had three or more, it was marked as high (green) likelihood of return on investment.

Based on the findings of this report summarized in Table 4, the highest return on investment technologies for fire management are first to develop single MAV Autonomy, Autonomous Sensing over fire, and the associated data and information system for active fire local environment mapping. Once this is completed for a single MAV, expanding the work to include many in a swarm would require further investment of distributed MAV autonomy and MAV swarm mechanics, but could greatly expand the breadth of application over large fires.

Important to investing in these technologies will be in collaborating with the key influencers and champions for using UAS technology in fire management. This includes working with UASAG, who plays a crucial role in how UAS technology is used by federal land management agencies, and with business entities already dedicated to technology transfer for fire management. Also, we advise working with NASA Ames that has already fostered a robust relationship UASAG and FAA. These three groups will help provide feedback of the utility of the technology, as well as facilitate integration into every day management operations.

Table 4: Synthesized findings for each use case in the context of investment for fire applications. Stakeholders are defined as the recipients of the technology. Green, yellow, and red denote high, medium, and low likelihood of return on investment. Impact is determined by Stakeholder Priority, Maturity is by Current TRL/ARL, and Feasibility is the change in TRL/ARL ("Delta TRL/ARL") after JPL Investment ($ Cost) and the potential funding sources, thus providing insight into the likelihood of outside funding after JPL investment. For Maturity, TRL/ARL 1-3 is considered good investment, 4-6 is a cautioned investment and 7-9 is a risky investment. Delta TRL/ARL is considered good investment if the technology increases in readiness by at least 3 levels, a cautious investment with increase of 2 levels, and a risky investment with an increase of 1 level. Feasibility of investment cost is partitioned based on RTD funding levels assuming an RTD Topical at $200k/yr for 3 years is a good investment, an SRTD at $600k/yr for 3 years is a cautionary investment, and anything more than that is risky. If there are 3+ potential future funding sources, an investment is considered good, 2 it is considered cautionary, and 1 it is considered risky. * denotes those recommended for investment.

| Use Case | Impact — Stakeholder Priority | Maturity — Current TRL/ARL | Feasibility — End TRL/ARL (delta TRL/ARL) | Feasibility — Investment Cost ($k) | Potential Funding Sources |
|---|---|---|---|---|---|
| *MAV Autonomy over fire | High | 2 | 6 (4) | Moderate | NASA ASP; NASA Planetary Flight; DARPA; ONR; NSF |
| Distributed MAV Autonomy | High | 2 | 6 (4) | Moderate | |
| *Autonomous Sensing over fire | High | 2 | 6 (4) | Low | |
| Swarm Mechanics for fire | Med | 2 | 6 (4) | Moderate | DARPA, ONR, ARMY |
| Data and Information Systems: Escape Routes | Low | 1 | 4 (3) | Low | NASA ASP; NASA AIST; CalFire; USFS; DHS (e.g., AUDREY) |
| Data and Information Systems: Person Location (no GPS) | Low | 3 | 6 (3) | Low | |
| Data and Information Systems: Person Geolocation (GPS) | High | 3 | 7 (4) | Low | |
| * Data and Information Systems: Active Fire Local Environment Mapping | High | 1 | 4 (3) | Low | |
| Data and Information Systems: Vegetation Stress | Low | 3 | 6 (3) | Low | |
| Data and Information Systems: Damage Assessment | High | 1 | 4 (3) | Low | |
| Hydro Propulsion | Low | 2 | 5 (3) | Low | Private Sector investors (Fire, Oil, Agriculture, Mining, etc.); NIST; USFS |
| Mobile, AI-deployed Infrasound Projectors | Low | 1 | 3 (2) | Moderate | NASA, ONR, DARPA, ARMY |

Applications from Space, in: Earth Science Applications from Space: 2017 Decadal Survey Request for Information. pp. 3–6.

Tyndall, J., 1874. Transparency and Opacity in the Atmosphere. Philos. Mag. J. Sci. 47, 374–384.

USFS, 2015. The Rising Cost of Wildfire Operations: Effects on the Forest Service's Non-Fire Work.

Westerling, A.L., 2006. Warming and Earlier Spring Increase Western U.S. Forest Wildfire Activity. Science (80-. ). 313, 940–943. doi:10.1126/science.1128834

Yun, K., Bustos, J., and Lu, T., 2018. Predicting Rapid Fire Growth (Flashover) Using Conditional Generative Adversarial Networks. Electronic Imaging. 127-1-127-4(4).
© 2018, California Institute of Technology. Government sponsorship acknowledged.    Page | 21